\documentclass[conference]{IEEEtran}
\usepackage[draft]{hyperref}
\usepackage[utf8]{inputenc}
\usepackage[english]{babel}
\usepackage{amsmath, amsfonts, amssymb, graphicx, microtype, float, url, textcomp, subcaption, tikz}
\usepackage[autolanguage]{numprint}
\usepackage[inline,shortlabels]{enumitem}

\newcommand{\orcidicon}{\includegraphics[width=0.32cm]{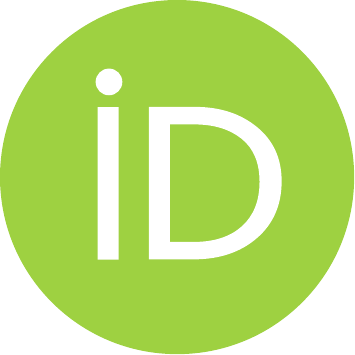}}

\usepackage[font=small]{caption}
\usepackage[capitalize]{cleveref}

\title{LemonLDAP::NG\\A Full AAA Free Open Source WebSSO Solution}

\foreach \x in {A, ..., Z}{%
	\expandafter\xdef\csname orcid\x\endcsname{\noexpand\href{\csname orcidauthor\x\endcsname}{\noexpand\orcidicon}}
}

\author{\IEEEauthorblockN{Christophe Maudoux\orcidA{}}
	\IEEEauthorblockA{
		\textit{CNAM/Cedric}\\
		Paris, France \\
	}
	\and
	\IEEEauthorblockN{Selma Boumerdassi\orcidB{}}
	\IEEEauthorblockA{
		\textit{CNAM/Cedric}\\
		Paris, France \\
	}
}

\begin{document}
	\maketitle

\begin{abstract}
Nowadays, security is becoming a major issue and concern. More and more organizations like hospitals, metropolis or banks are under cyberattacks and have to improve their network infrastructure security.
The first prerequisites are to authenticate users, to provide identity and to grant just the needed and useful accesses. These requirements can be solved by implementing a Single Sign-On (SSO) solution. It is an authentication scheme that permits a user to log in with a single identity to any of several related, yet independent, systems. It  allows users to log in once and to access services without authenticating again. SSO solutions are classified depending on \emph{Authentication}, \emph{Authorization}, and \emph{Accounting} features. The 'AAA' acronym defines a framework for intelligently controlling access to resources, enforcing security policies, auditing usage, and providing the information necessary to bill for services. These combined processes are considered \emph{important} for effective \emph{network management} and \emph{cybersecurity}. \texttt{LemonLDAP::NG} (LL::NG) is a full AAA WebSSO solution. It implements all standard authentication and identity federation (IdF) protocols. The main LL::NG's advantages compared to other products are its plug-in engine and its advanced handler-based protection mechanism that can be employed to protect Server2Server exchanges or to offer the SSO as a Service, a solution to implement a full DevOps architecture. LL::NG is a community and professional project mainly employed by the French government to secure Police, Finance or Justice Ministries and a French mobile operator IT infrastructures since 2010. But for several years, contributions come from all around the world and LL::NG is becoming more and more popular.
\end{abstract}



\section{Introduction}
SSO is a centralized session and user authentication service in which one set of login credentials can be used to access multiple applications. Main advantage is in its simplicity; the service authenticates you on one designated platform or portal, enabling you to then use a variety of services without having to log in and out each time. Deploy an SSO architecture avoids multiplication of passwords and increase the overall IT security. Full SSO solutions provide the three essential services that are users authentication with different methods, access control with authorizations, and accounting by providing logs.

\begin{figure}
	\centering
	\includegraphics[width=\linewidth, height=150px]{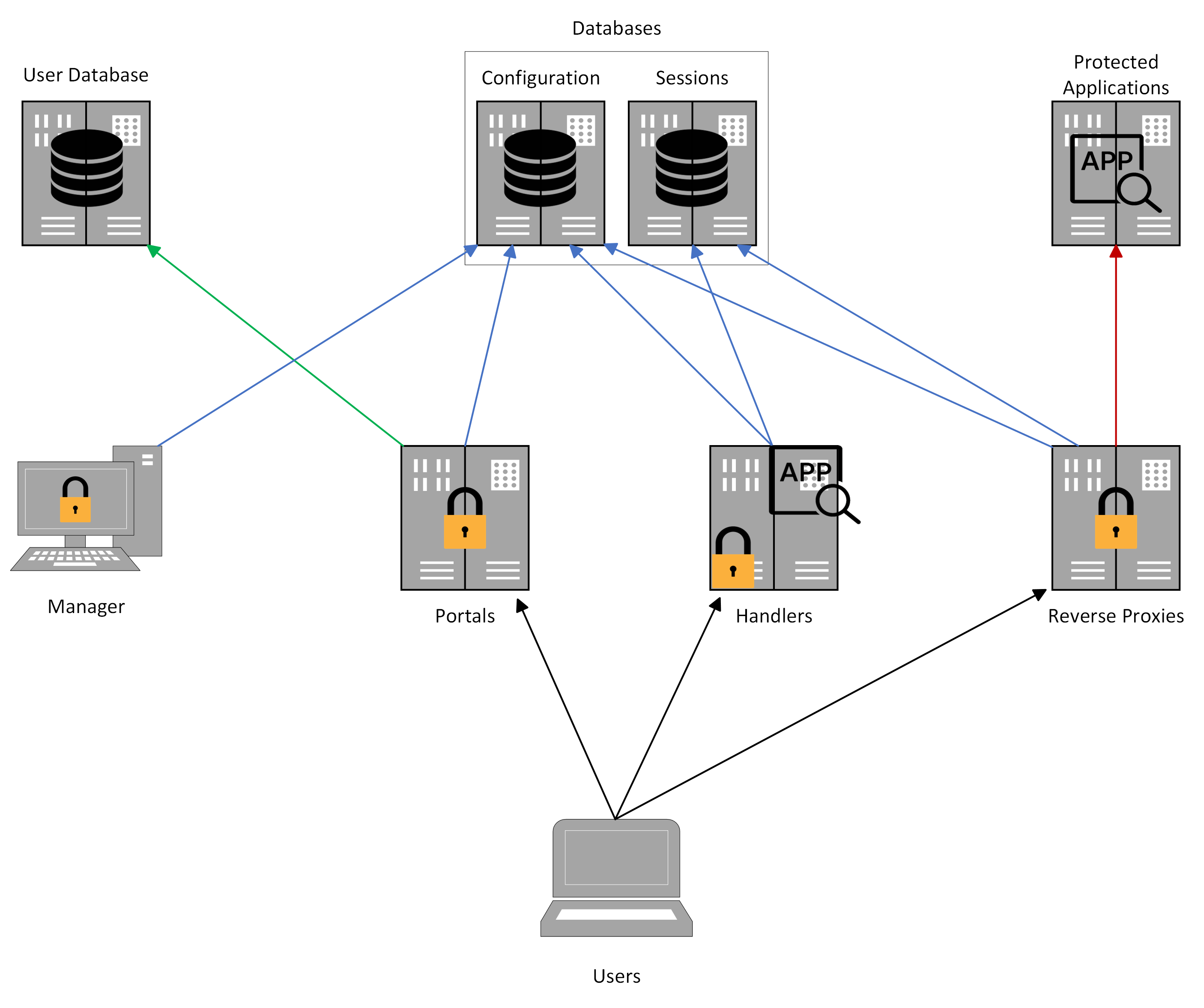}
	\caption{Main components}\label{fig:archi.png}
\end{figure}

\texttt{LemonLDAP::NG} \cite{guimardLemonLDAPNG2010} works with PSGI, FastCGI or uwsgi \cite{UWSGIProjectUWSGI} standards/protocols compliant web servers like \texttt{Nginx}, \texttt{Apache}, \texttt{Starman}. It offers different ways for \emph{authenticating} users, checks \emph{authorizations} before accessing resources, and provides logs for \emph{accounting}. It is composed of four main components as depicted by \cref{fig:archi.png}.

\begin{enumerate*}[\itshape(i)]
	\item The \emph{Portal} implements standard protocols. When a user tries to access an application protected by LL::NG and connected by using authentication or IdF protocols, he is redirected to the \emph{Portal}. It displays authentication screen and then, redirects user to the requested application or it displays the applications menu.
	\item The \emph{Handler} (depicted by a lock) can be employed to protect applications that do not implement standard authentication or IdF protocols but working with HTTP headers. Handler code can be embedded by applications or Reverse Proxies as explained in \cref{subsec:handlers}. When an unauthenticated user tries to access an application protected by a LL::NG handler, he is redirected to the Portal by the Handler. Then, the Portal displays authentication screen, builds a SSO session in the sessions back-end, provides a SSO cookie with the Session ID (SID) and redirects the authenticated user to the requested application. With the SID, the Handler can retrieve sessions data and send session attributes to the protected application by using HTTP headers.
	\item LL::NG relies on different \emph{back-ends} that can be files, LDAP, Active Directory, SQL or noSQL (for SSO sessions) databases to store configuration, SSO sessions and persistent sessions (with history, second factors, consents), and users data.
	\item The \emph{Manager} is the WebSSO administration interface used by administrator to set access rules, HTTP headers sent to protected applications, define the applications menu, configure identities and services providers, enable authentication methods.
\end{enumerate*}

Rest of this paper is organized as follow. \Cref{sec:background} is an overall presentation of the LL::NG project. In \cref{sec:auth-methods}, we expose main authentication methods. Standard authentication or IdF protocols and applications protection are described in \cref{sec:appprotection}. LL::NG implements a plug-ins engine and some community supported plug-ins are presented in \cref{sec:plugins}.

\section{Background \& Related Work}\label{sec:background}
SSO platforms can be implemented for different use cases. To increase security, we can not rely on heterogeneous products. So, it is important to provide complete solutions like LemonLDAP::NG.
\subsection{History \& Development}
The first stable LL::NG version has been released in 2010. It was a fork of \texttt{LemonLDAP}, an OpenLDAP-based authentication project that has been modified by the French 'Gendarmerie Nationale' (GN) to meet its specific needs.

Since 2004, GN is part of the LL::NG core team and a major contributor to this open source project. In 2016, the 1.9 release provides a new \textsc{AngularJS} Manager and supports the OpenId Connect protocol. The fully re-coded 2.0 version, released in 2018, provides a new plug-ins engine and supports Multi Factor Authentication (MFA) \cite{maudouxImplementationAuthentificationDouble}. LL::NG 2.0.15-2, the last stable version that offers several new features, can be downloaded since September, 2022 from the OW2 official website \cite{OW2ProjectsLemonLDAP}.

LL::NG is mainly coded in \texttt{Perl} and \texttt{JavaScript} by the core development team members: \textsc{Xavier Guimard} (GN/STSISI), \textsc{Christophe Maudoux} (GN/STSISI \& Cnam), \textsc{Clément Oudot} (Worteks) and \textsc{Maxime Besson} (Worteks).
'Worteks' \cite{worteksWorteksExpertiseOpen} is an expertise and publishing company in free and open source software. The 'STSISI' department's primary mission is to manage the technological modernization of the French 'Police Nationale' and 'Gendarmerie Nationale'.

\begin{figure}[b]
	\centering
	\includegraphics[width=.85\linewidth]{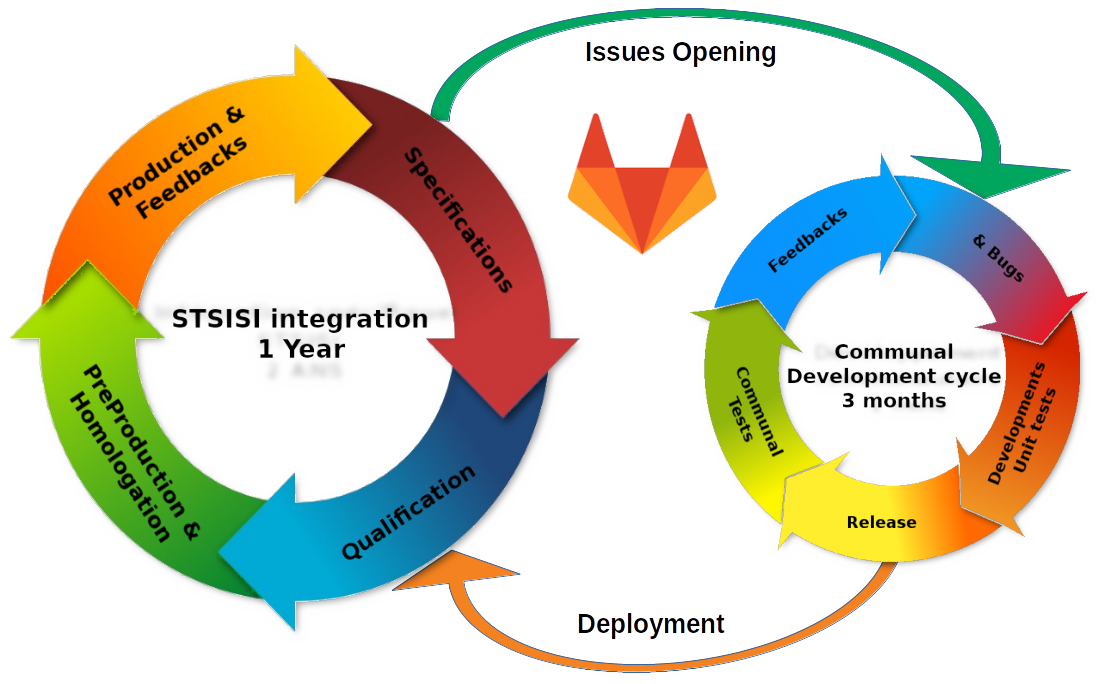}
	\caption{Development process}\label{fig:cycle.png}
\end{figure}

\Cref{fig:cycle.png} describes the LL::NG development process that is composed of two different cycles.
\begin{enumerate*}[\itshape(i)]
	\item The community cycle, hosted by the OW2 GitLab forge \cite{GitLabOW2LemonLDAP}, is used for opening issues, Continuous Integration \& Development (CI/CD) pipelines and milestones.
	\item Then, when a new version is released by the community every three months, this one is tested, validated, and deployed during the STSISI 1-year cycle.
\end{enumerate*}

\subsection{Related Work}
Some SSO solutions have been already implemented to protect data collection and server to server exchanges. Study in \cite{chatterjeeApplyingSpringSecurity2022} describes how \texttt{Spring Security} and the \texttt{Keycloak} open-access platform can be employed to secure a microservices-based architecture Application Programming Interfaces (APIs). Embedded protection uses the \texttt{OAuth2} method, the underlying concept behind the OIDC protocol that is presented in \cref{subsec:oidc}. Problem here is that the \texttt{OAuth2} solution only consists in authenticating users. It can not be used for providing user's identity attributes to services unlike the OIDC protocol.

Some works have been conducted in \cite{dsilvaBuildingZeroTrust2021a} to deploy a Zero Trust Architecture (ZTA) using \texttt{Kubernetes} and a SSO product. As explained by the authors, a ZTA keeps checking the user’s authenticity, monitors the user’s devices, and checks for any location change initiated by the user device. Moreover, it also regularly checks for any discrepancies in the application that the user would be using. Main drawback here compared to \emph{ServiceToken} handler explained in \cref{subsec:server2server} is that none authenticated logs are generated to track users activities.


\section{Users Authentication}\label{sec:auth-methods}
Many protocols or methods can be employed or combined to authenticate users. This section provides an overview of the main ones supported by LemonLDAP::NG.
\subsection{Methods \& Protocols}
LL::NG provides several authentication methods and implements standard protocols. Users can authenticate by using network protocols like \emph{Gnu Private Guard} (GPG) \cite{IndexGnuPGWiki} that consists in asking user to sign a challenge and to post the result. \emph{Remote Authentication Dial-In User Service} (Radius) is a client-server networking protocol that runs in the application layer \cite{HowDoesRADIUS}. LL::NG handles and forwards Radius authentication requests to the Radius server. With \emph{Kerberos} \cite{KerberosNetworkAuthentication}, users are authenticated based on their desktop session. LL::NG validates the Kerberos ticket against a \emph{local keytab}. LL::NG can also rely on an \emph{LDAP} directory \cite{OpenLDAPMainPage} to authenticate users and to get user attributes. LL::NG is compliant with LDAP v2 or v3 server, including \emph{Active Directory} (AD) \cite{ActiveDirectory}, and is compatible with LDAP password policy to check password strength, to block brute-force attacks or can force password change on first connection. LL::NG proposes \emph{Pluggable Authentication Module} (PAM) as a simple authentication back-end. It is a mechanism to integrate multiple low-level authentication schemes into a high-level Application Programming Interface (API). PAM permits software that rely on authentication to be written independently of the underlying authentication scheme.

Connectors based on \emph{OAuth2} protocol \cite{OAuthOAuth} are available like \emph{Google}, \emph{GitHub}, \emph{Facebook}, \emph{LinkedIn} or \emph{Twitter} that allows applications to reuse its own authentication process \cite{dodanduwaRoleTrustOAuth2018}.

Main authentication and standard SSO protocols like \emph{CAS} or \emph{SAMLv2} and \emph{OIDC} described in \cref{sec:appprotection} are also implemented for authentication, authentication delegation or identities federation \cite{setteIntegratingCloudPlatforms2014}.

Furthermore, “combo” back-ends like \emph{Authentication choice} and \emph{Combination} can be employed to propose different authentication methods or to combine sequences.

\subsection{Multi Factor Authentication}\label{subsec:mfa}
MFA is supported since the LL::NG 2.0 release. This feature can be activated for confirming a user’s claimed identity by using a combination of two different factors \cite{huseynovChapter50ContextAware2017} between:
\begin{itemize}
	\item something you \emph{know} (login/password, \dots)
	\item something you \emph{have} (U2F Key, TOTP, mail, \dots)
	\item something you \emph{are} (biometrics like fingerprints, \dots)
\end{itemize}

LL::NG provides some second factor plug-ins that can be enabled for completing authentication module with MFA:
\begin{description}
	\item[TOTP] Time based One-Time Password is an algorithm that computes a one-time password from a shared secret key and current time \cite{WhatAreTime}. Then, users might register a device as smartphone by using a dedicated application like \textsc{FreeOTP} or \textsc{Google Authenticator}
	\item[U2F] Universal 2nd Factor is an open authentication standard that enforces and simplifies two-factor authentication using specialized USB or NFC devices \cite{WhatFIDOU2F}. LL::NG can propose to users to register their key(s). Then, 2F registered users can not login without using their key(s)
	\item[Yubikey] It is a hardware token manufactured by 'Yubico'. It sends an OTP, which is validated via Yubico server \cite{LetGetStarted}
	\item[Mail] After logging in via an authentication module, a one-time code is generated by the Portal and sent to the user’s e-mail address. Then, user is prompted for this code in order to continue and to achieve the login process
	\item[External] This method can be used to append a second factor authentication device like SMS, OTP, and so on. It calls external commands to send or to validate a second factor
	\item[REST/Radius] Those methods rely on the corresponding protocol to submit and to check the second factor
\end{description}


\section{Applications Protection}\label{sec:appprotection}
Once users are authenticated, identity claims might be sent or provided to protected applications. Depending on applications and their capabilities different solutions can be used.
\subsection{Central Authentication Service}
CAS is an enterprise SSO solution and identity provider for web applications \cite{CASHome,huApplicationCrossdomainSingle2013}. It is an open and well-documented authentication protocol. When a client try to access a “CAS-sified” application which is an application requiring CAS authentication, the application redirects it to the CAS server. 
CAS validates the client's authenticity, usually by checking a username and password against a database.

Unlike protocols below, CAS requires to modify the application code to authenticate and retrieve the user data.
\subsection{Security Assertion Markup Language version 2}
SAMLv2/Shibboleth is an open standard used for authentication \cite{SAMLExplainedPlaina,sobhIdentityManagementUsing2019}. Based upon the Extensible Markup Language (XML) format, web applications rely on SAML to transfer authentication data by using assertions (authentication request and response) between two parties -- the Identity Provider (IdP) and an application known as the Service Provider (SP). 
SAMLv2 protocol requires an approbation link between the IdP and the SP. In fact, IdP and SP might exchange their respective metadata that permits to provide configuration parameters like end points and public keys.

SAMLv2 protocol does not require a direct link between IdP and SP. All network exchanges go through the browser that is not the case with OIDC.

\subsection{OpenID Connect}\label{subsec:oidc}
OIDC is an evolution of SAMLv2 \cite{OpenIDConnectExplained}. It employs REST protocol instead of SOAP to exchange assertions and JSON replaces XML to format messages. OIDC is easier to implement with mobile applications by using the refresh token mechanism.

OIDC standard defines 3 authentication flows. The \emph{Hybrid} and \emph{Implicit} flows are \emph{deprecated} and should not be employed. The \emph{Authorization Code} (AzC) flow is the most secured because user and applications are authenticated. The AzC is returned by the OP after user authentication. This code is provided to the application. This flow is used to prevent that the password is stolen by a malicious application. Then, it is provided with the \emph{Client Id} and \emph{Secret} to authenticate the application, and to retrieve \emph{Id}, \emph{Access} and \emph{Refresh} tokens.

LL::NG implements all these standard SSO protocols. It can act as IdP/SP (SAML) or OP/RP (OIDC) and could be used as a “federation proxy” between all these protocols \cite{LLNGFederation}.

\subsection{Handlers}\label{subsec:handlers}
Protection by handlers is a mechanism that can be employed to protect applications working with HTTP headers. Each request sent to the protected application is caught by the handler. It looks for a specific “token” depending on the used handler type. LL::NG proposes the following types:
\begin{description}
	\item[Main] SID is retrieved from the SSO cookie value
	\item[AuthBasic] Performs Basic authentication \cite{IBMDocumentation2021}
	\item[SecureToken] Decrypts SID from a ciphered SSO cookie
	\item[ServiceToken] Retrieves SID from \emph{X-LLNG-TOKEN} header
	\item[DevOps] Fetches configuration from 'rules.json' file
	\item[OAuth2] Gets SID from an OAuth2 Access token
	\item[Zimbra] Performs Zimbra pre-authentication \cite{PreauthZimbraTech}
	\item[CDA] Used for Cross Domain Authentication \cite{huApplicationCrossdomainSingle2013,WhatAreCross2018}
\end{description}

With the SID, handler collects user data from SSO sessions back-end, computes and checks access rules and sends HTTP headers with session attributes to the protected application.

Handler code can be embedded by the protected applications but it requires to have access to the application code and to modify it.	To avoid this, applications can be hidden behind Reverse Proxies that embed the handler code. So, protected applications are not exposed to users and directly requested. Reverse Proxies-based architecture is described by \cref{fig:archi.png}.

\Cref{fig:handler.png} details the Main handler kinematic:
\begin{figure}
	\centering
	\includegraphics[width=\linewidth]{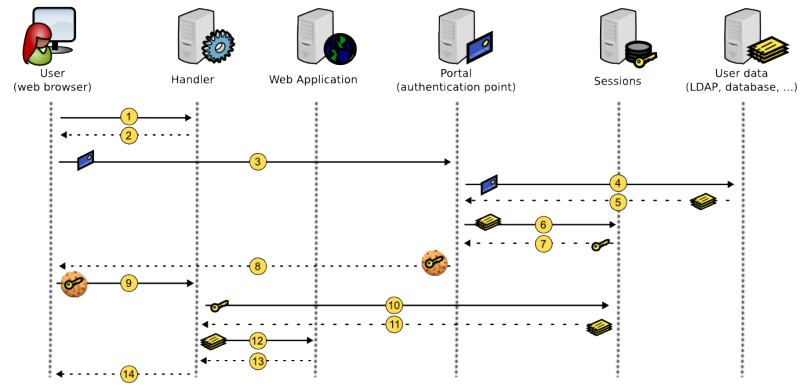}
	\caption{Main Handler kinematic}\label{fig:handler.png}
\end{figure}
\begin{enumerate*}
	\item User tries to access a protected application. His request is caught by Handler
	\item SSO cookie is not detected, so Handler redirects user to Portal
	\item User authenticates on Portal
	\item Portal checks authentication
	\item If authentication succeeds, Portal collects user data
	\item Portal creates a session to store user data
	\item Portal gets the SID
	\item Portal creates a SSO cookie only valid on the SSO domain with the SID as value
	\item User is redirected on the protected application with a SSO cookie
	\item Handler gets SID from the cookie and retrieves user session data
	\item Handler stores user data in its cache
	\item Handler checks access rules and sends HTTP headers to the protected application
	\item Protected application sends response to Handler
	\item Handler forwards the response to user.
\end{enumerate*}

\section{Extended Features}\label{sec:plugins}
LemonLDAP::NG offers possibility to extend its features by hooking internal processes or appending functionalities.
\subsection{Plug-ins Engine \& Entry Points}
Since 2.0 version, LL::NG provides a plug-ins engine. It allows to append specific or custom features. Plug-ins can handle authenticated or unauthenticated routes. They can also be launched by using particular entry points (EP) during login or logout process. This mechanism is a very interesting feature to extend LL::NG and to meet specific needs.
%
%

\subsection{Specific Plug-ins}
\begin{description}
	\item[\texttt{AdaptativeAuthLevel}] A user reaches an authentication level depending on the used authentication module, and eventually MFA as explained in \cref{subsec:mfa}. This plug-in adapts the authentication level depending on other conditions, like network or IP address, device, and so on
	\item[\texttt{BruteForceProtection}] To prevent brute force attack by blocking account after several login failures
	\item[\texttt{CheckDevOps}] To validate 'rules.json' files (\cref{sec:plugins})
	\item[\texttt{CheckUser}] To check session attributes, access rights and transmitted headers for a particular user and URL
	\item[\texttt{ContextSwitching}] Specific users like 'admins' can switch context other user for debugging purpose. Beginning and end of context switching process are logged
	\item[\texttt{CrowdSec}] A free and open-source security automation tool leveraging local IP behaviour detection and a community-powered IP reputation system \cite{FirewallBouncerCrowdSec}
	\item[\texttt{Impersonation}] Users can assume identity of another user for training or teaching purpose
	\item[\texttt{Notification}] To notify messages to users when log in
	\item[\texttt{REST/SOAP}] To provide the corresponding services
\end{description}

\section{Advanced Usages}\label{sec:advanced-usages}
LemonLDAP::NG implements all the previous features. But some specific needs can only be met by using advanced functionalities or mechanisms exposed below.
\subsection{Server2Server Exchanges Protection}\label{subsec:server2server}
\begin{figure}
	\centering
	\includegraphics[width=\linewidth,height=140px]{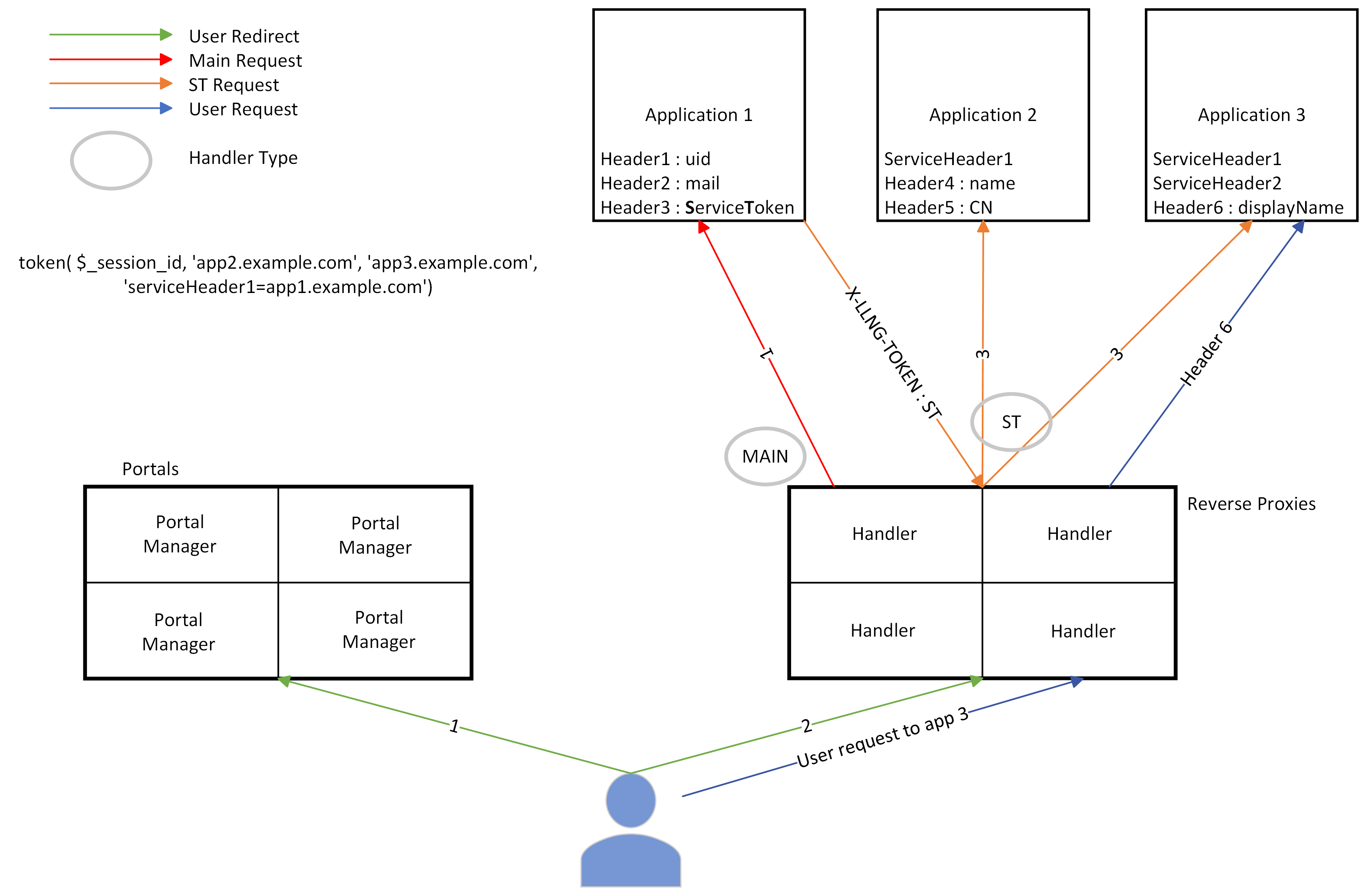}
	\caption{ServiceToken handler kinematic}\label{fig:servicetoken.png}
\end{figure}

The \emph{ServiceToken} handler is a mechanism to protect \emph{Server2Server exchanges}. As described by \cref{fig:servicetoken.png}, a web application (\texttt{WebApp1}) may need to request some other web applications (\texttt{WebAppN}) on behalf of the authenticated user. You just have to provide a header containing the \emph{ServiceToken} (ST) to \texttt{WebApp1} and to protect \texttt{WebAppN} with the corresponding handler. Then \texttt{WebApp1} can request \texttt{WebAppN} by appending the 'X-LLNG-TOKEN' HTTP header with the ST as value that will be caught by the handler.

The ST is built by encrypting the SID and an authorized VHosts list to restricted its scope. The ST lifetime is limited in case of it is stolen. The ST can also be used for sending service headers to \texttt{WebAppN} like the source host by example.

\subsection{SSO as a Service}
The \emph{DevOps} handler is a mechanism to implement the \emph{SSO as a Service} (SSOaaS). '*aaS' means that application can drive underlying layer like IaaS for Infrastructure or PaaS for Platform. SSOaaS means provide the ability for an application to manage authorizations and choose user attributes to set. But authentication can not be really *aaS. This step must be performed by a well known, identified and managed service. It can not be delegated except with IdF protocols as described and explained in \cref{sec:appprotection}. Applications might just use it but not manage it. LL::NG offers some features that can be used for providing SSOaaS and implementing DevOps architectures.

DevOps concept means “development” and “operations”. It is the combination of practices and tools designed to increase an organization’s ability to deliver applications and services faster than traditional software development processes. This speed enables organizations agility to better serve their customers and compete more effectively in the market. DevOps breaks barriers between traditionally isolated development and operations teams. Under a DevOps model, development and operations teams work together across the entire software application life cycle, from development and test through deployment to operations. It means in a SSO architecture that a web application can manage its own rules and headers regardless the SSO administration team.

\begin{figure}
	\centering
	\includegraphics[width=\linewidth,height=160px]{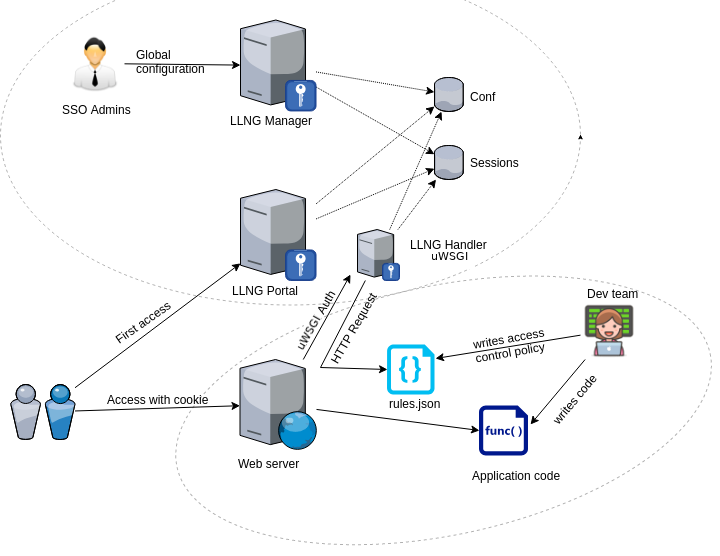}
	\caption{DevOps handler kinematic}\label{fig:devops.png}
\end{figure}

The DevOps handler is an on-premises component (the LL::NG handler) designed to retrieve users data or SSO configuration from back-ends and VHost configuration (access rules and sent HTTP headers) not from LL:NG configuration but from the web application itself that can be hosted in a cloud environment. Rules and headers are set in a 'rules.json' file stored at the website root directory as described by \cref{fig:devops.png}. Communications between DevOps handler and web applications are based on a Common Gateway Interface (CGI) protocol like FastCGI or uwsgi. Rules.json files syntax and format can be checked by the DevOps teams with the \texttt{CheckDevOps} plug-in described in \cref{sec:plugins}.

\section{Conclusion}\label{sec:conclusion}

\textsc{LemonLDAP::NG} \cite{LemonLDAPNGWeb} is a free and open source full AAA SSO project that supports standard SSO protocols (SAMLv2, CAS, OIDC, \dots), provides several features (MFA, different handlers or authentication methods) and services. It is a complete, versatile and customizable SSO solution. Its main specific and original features compared to the few other existing products are SSOaaS and Server2Server protection mechanisms. Thus and standard IdF protocols can be deployed and implemented in a cloud infrastructure to easily authenticate users, protect web applications and provide logs. It can also be employed as a proxy to interconnect different cloud platforms or between heterogeneous identity federation protocols.

Next release will offer a new Manager based on \texttt{ReactJS} framework, implement the \texttt{WebAuthn} protocol and a new MFA engine.
	\bibliographystyle{IEEEtran}
	\bibliography{IEEEabrv,llng}
\end{document}